# The Josehedron

## A space-filling plesiohedron based on the Fischer-Koch S Triply Periodic Minimal Surface

Mathias Bernhard, ETH Zurich, bernhard@arch.ethz.ch

**Abstract**: This paper presents a novel space-filling polyhedron (SFPH), here named the *Josehedron*, derived from the extremal points of the Fischer-Koch S triply periodic minimal surface (TPMS). The *Josehedron* is a plesiohedron with 12 faces (4 isosceles triangles and 8 mirror-symmetric quadrilaterals), 12 vertices, and 22 edges. It tiles three-dimensional space with 12 instances per cubic unit cell in 6 distinct orientations. The generating point set exhibits a remarkable connection to the pentagonal Cairo tiling when projected onto any coordinate plane. Several additional geometric properties are described, including integer vertex coordinates, interwoven labyrinths, and chiral symmetry between the polyhedra obtained from the combined minima and maxima of the function. Finally, the paper presents a general method for finding novel SFPHs based on any periodic function, TPMS, or other functions. The described method is applied to a selection of TPMS, and 7 additional, previously undocumented SFPH are shown in the Appendix.

# 1. Introduction

## 1.1. Space-filling polyhedra in general

The study of space-filling polyhedra (SFPH), also known as honeycomb cells or tessellation polyhedra, is a long-established yet still active area of research[1,2], with applications spanning crystallography, pure mathematics, architecture, and materials science[3]. A space-filling polyhedron is one whose congruent copies can tile all of three-dimensional Euclidean space without gaps or overlaps, through some combination of translations and rotations. If reflections (mirror copies of the original SFPH) are also involved, the term “chiral” is used.

Of the five Platonic solids – even though Aristotle made the math world believe otherwise for over 18 centuries[4] – only the cube possesses this property. Among the Archimedean solids and other well-known polyhedra, several additional space-fillers have been identified. Some of these are composed of two types of regular polygons, such as the truncated octahedron, the gyrobifastigium, and the triangular and hexagonal prisms, while others have identical faces, including the rhombic dodecahedron and the flat octahedra formed by connecting cube face centers to cube vertices[5,6].

Space-filling polyhedra have long captivated both professional mathematicians and amateur enthusiasts alike[7,8]. Beyond the classic examples, other irregular space-filling types have been discovered. Notably, Grünbaum and Shephard presented further examples in the late 1970s, demonstrating that the landscape of space-filling polyhedra is richer and more varied than previously assumed. New shapes keep being discovered in nature[9] or invented, refuting old conjectures[10]. In the literature prior to 1980, a total of sixteen dodecahedra (12 faces) were known to fill space[11]. This paper introduces yet another such polyhedron.

## 1.2. Plesiohedra

A particularly important family of space-filling polyhedra is the plesiohedra. A plesiohedron is defined as the Voronoi cell of a point in a discrete point set: the region of space that is closer to that point than to any other point in the set. By construction, the Voronoi cells of any discrete point set tile space without gaps or overlaps, making every plesiohedron a space-filler. Plesiohedra are called «Wirkungsbereichs-polyedertypen» by Engel[12].

Space-filling polyhedra can be organized in a hierarchy of increasingly general families. The most restrictive are the parallelohedra, of which there are exactly five types, and which tile space by translation alone. Parallelohedra form a subset of plesiohedra, which in turn are a subset of stereohedra (polyhedra that tile space using translation, rotation, and reflection). Stereohedra, finally, are a subset of all space-filling polyhedra.

Several well-known triply periodic minimal surfaces (TPMS) give rise to plesiohedra through this Voronoi construction. For instance, the extremal points of the Schwarz P surface generate cubes, while those of the Gyroid surface, whose skeleton is the Laves graph, yield a 17-sided plesiohedron. Some plesiohedra are even shown to have remarkable similarities to TPMS themselves, not in the kind of dual relationship via the skeleton described above[13]. The present paper applies the same approach to the Fischer-Koch S (FKS) surface to obtain a new 12-faced plesiohedron. An overview of the resulting polyhedra resulting from this construction is given in Table 1.

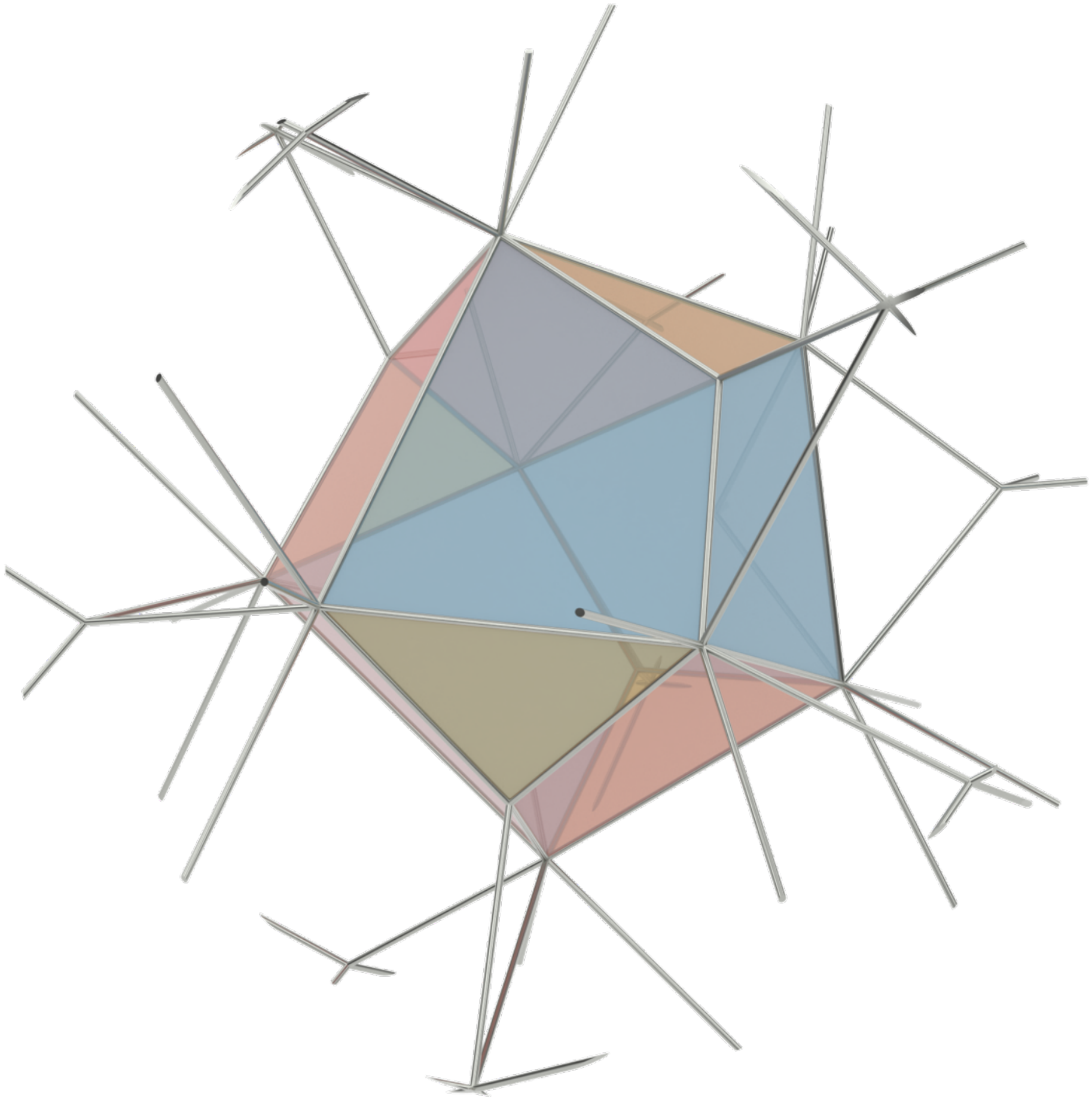

*Fig. 1: The Josehedron, with edges of neighboring cells, faces colored by type*

# 2. Method

The method used in this paper consists of identifying the extremal points (minima and/or maxima) of a triply periodic minimal surface function, and then constructing their Voronoi tessellation to obtain the plesiohedra. a special case of stereohedra[14,15].

*Table 1: Overview of some TPMS and their corresponding SFPH*

| Name | space filling polyhedron | Range (*=rounded) |
|---|---|---|
| Diamond | Min: Triakis truncated tetrahedron<br>Max: same<br>Both: truncated octahedron | [-√2, √2] |
| Double Diamond | Min: does not apply, just a shell around the D surface<br>Max: truncated octahedron<br>Both: does not apply | [-1, 3] |
| Double Gyroid | Min: truncated octahedron<br>Max: same as both of Gyroid<br>Both: **New type**, 20 F, two big parallel faces, see Fig. 12 b | [-3, 4.125]* |
| DP | Min: cube<br>Max: truncated octahedron<br>Both: cubes with chamfered vertices and octahedra | [-0.600, 2.099]* |
| Fischer Koch S, FKS | Min: ***Josehedron***, described in detail below<br>Max: ***Josehedron***, mirrored<br>Both: **New type**, see Fig. 6 | [-√2, √2] |
| FRP | Min: rhombic dodecahedron<br>Max: same<br>Both: **New type**, 20 F, perspectival cube with hexagonal corner, see Fig. 12 c | [-7.933, 9.460]*<br>[-7.863, 9.453]* (new) |
| Gyroid | Min: Plesiohedron, 17 F, Voronoi of Laves graph<br>Max: same<br>Both: **New type**, 17 F, two parallel equilateral hexagons, see Fig. 12 a | [-1.5, 1.5] |
| IWP | Min: cube with pyramids, 12 F, 4 x square, 8 x isosceles triangle<br>Max: truncated octahedron<br>Both: cube | [-5, 3] |
| KP | Min: flat octahedra, two square pyramids<br>Max: square pyramid, h=s/2 (cube center)<br>Both: **2 types**, single side truncated flat octahedron + capsule with rhombic tips, see Fig. 12 g | [-3.8, 6.6]* |
| Lidinoid | Min: truncated octahedron<br>Max: **New type**, 14 F, 2 truncated pyramids with chamfered edges, see Fig. 12 d<br>Both: **Various new types**, (to be verified if they are distinct or due to numerical imprecision) | [-1.35, 1.244]* |
| Neovius | Min: cube<br>Max: cube<br>Both: truncated octahedron | [-13, 13] |
| Octo | Max: truncated octahedron<br>Min: flat octahedra (cube face to center)<br>Both: 3 types; cube, rhombic dodecahedra, and 14 F polyhedra | [-6.3, 13.7]* |
| Schwarz P | Min: cube<br>Max: cube<br>Both: truncated octahedron | [-3, 3] |
| Split-P | Min: cube<br>Max: **New type**, truncated hexagonal pyramid with chamfered edges, 17 F, see Fig. 12 e<br>Both: **2 types**, 17 F and 20 F, see Fig. 12 f | [-1.800,1.801]* |

The FKS surface (also referred to as Fischer-Koch, FK-S or simply S) has various trigonometric approximations in the literature[16–20]. The formula used throughout this paper is:

$$cos(2x) * sin(y) * cos(z) + cos(2y) * sin(z) * cos(x) + cos(2z) * sin(x) * cos(y)$$

The formulae for the other TPMS types investigated in this study are shown in Table 6 in the Appendix. The surface was discovered by Werner Fischer and Elke Koch[21] in the late 80s. They did not describe the trigonometric approximation in the equation above, but demonstrated it with physical models made of sticks and cloth. They mention other approximations called “balance surfaces” based on the interpenetrating skeletons (see Fig. 9), also sometimes referred to as labyrinths[22–24]. This same strategy was used by the author to derive a polyhedral surface topologically equivalent to the diamond TPMS, but entirely composed of planar quads to improve printability via large-scale concrete extrusion[25,26].

The extremal points of the TPMS function were located by progressively adding a constant to the function value (effectively shifting the iso-surface threshold) until the iso-surface collapses to isolated points. While this thinning procedure does not theoretically guarantee convergence to the true skeleton nodes, it proved effective in practice for the surfaces examined here. The computation and geometric modelling were performed using the Rhino Grasshopper plugin Axolotl[27], developed by the author.

A less empirical, purely analytical method using numerical procedures such as `.minimize` or `.basinhopping` from the package `scipy.optimize` to find the minima and maxima of any function, trigonometric or otherwise, will be rigorously employed for the remaining candidate TPMS functions. The function `scipy.optimize.differential_evolution` can find global minima with very high precision. The `minimize` function found the values `3.1416333`, `3.14161617`, and `3.14158459` for a sample function. This is close to the actual value of π, but would need explicit knowledge and confidence to snap to it automatically.

A TPMS is conventionally defined as the zero-level iso-contour of a scalar field, where the Gaussian curvature is 0 everywhere on the surface. By adding a constant to the function value before extracting the iso-surface, one obtains a family of nested offset shells, analogous to the layers of an onion (Fig. 2). These are not true parallel offsets, since the trigonometric approximations used here are not metric signed distance functions. Nevertheless, progressively increasing the constant causes the iso-surface to shrink until it eventually degenerates into isolated points at the function extrema. For the FKS surface, these extreme values are $\pm\sqrt{2}$ and they occur at the locations listed in Table 4.

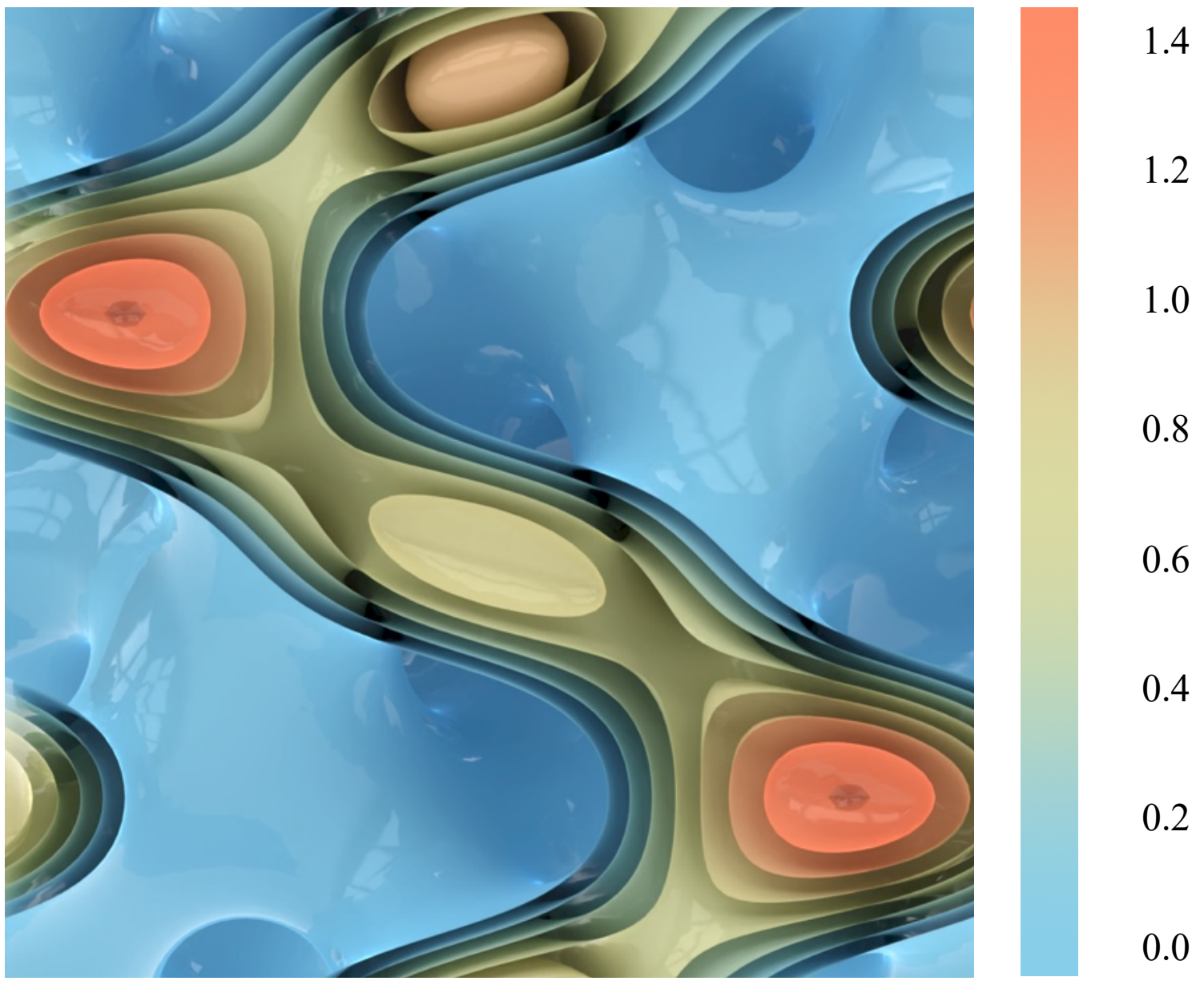

*Fig. 2: 8 onion shells at 0.2 intervals, from 0 (blue) to 1.4 (red)*

While ongoing research seeks to establish an upper bound on the number of faces for Dirichlet stereohedra[2], this paper presents one with only 12 faces. The shape itself and its spatial arrangement into a space-filling tessellation exhibit a plethora of interesting geometric properties, as described in the following section.

# 3. Result

## 3.1. Shape

Using the method described above, plesiohedra can in principle be derived from any TPMS. Some of these will have single congruent shapes, while others may come in pairs of chiral copies (mirror images of each other). The plesiohedron presented in this paper occupies a special place among these due to its high degree of regularity and numerous remarkable geometric properties. Naming policies for new shapes appear to be rather tolerant (looking at you, gyrobifastigium), so I decided to name the here presented SFPH the *Josehedron* (Fig. 1), after my daughter Josefine.

A notable property of the *Josehedron* is that all its vertex coordinates, when appropriately scaled, are integers. The squared distances from the origin to the vertices fall into two groups: eight vertices on the outer circumsphere lie at a squared distance whose coordinate components sum to 69, with two distinct triplets: 1, 2, 8 (Group 2) and 2, 4, 7 (Group 1), while four vertices, the flat apices of the triangular faces, lie on the inner circumsphere at a squared distance of 54, with the triplet 3, 3, 6 (Group 3, see Fig. 3).

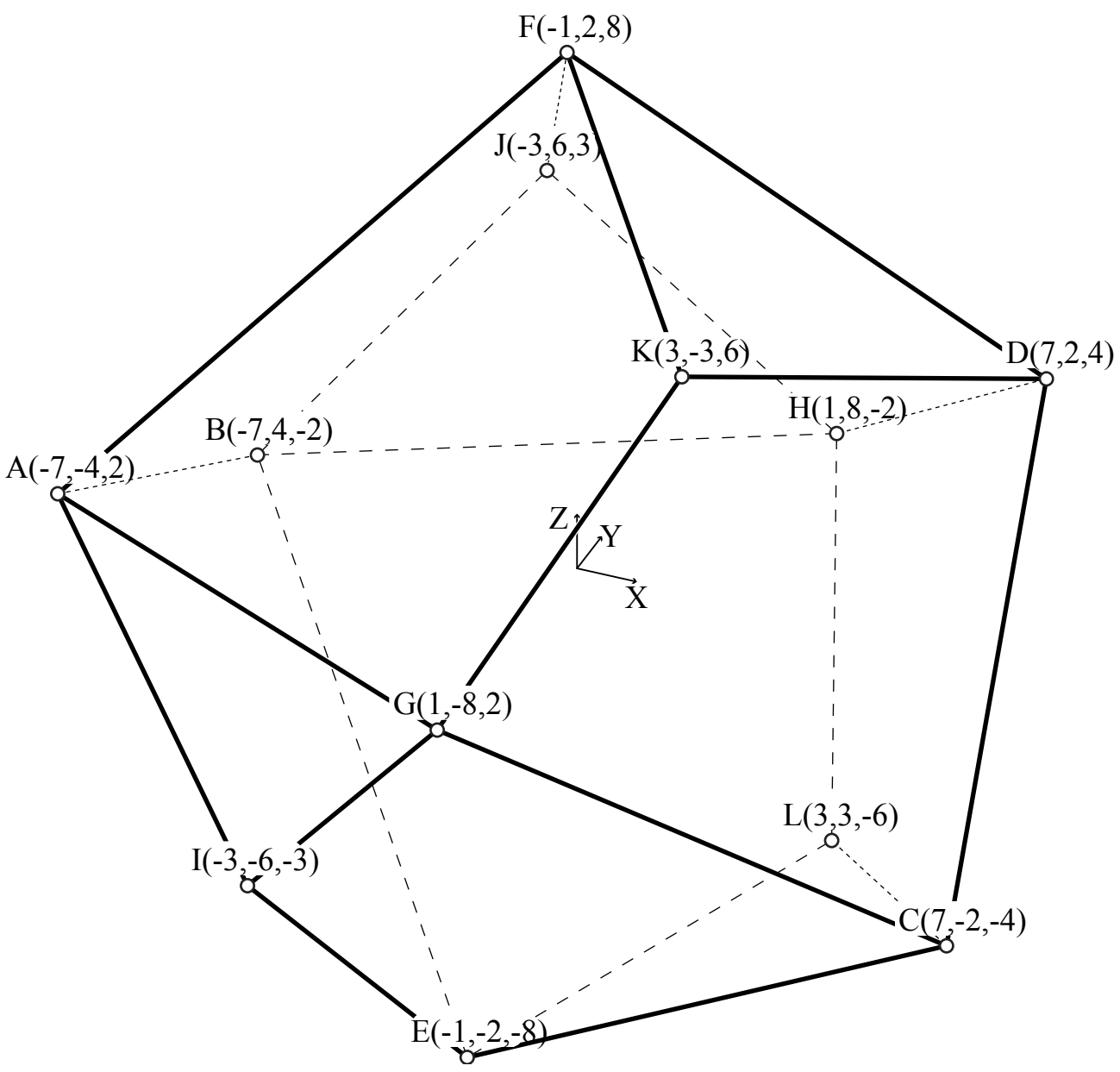

| Ortho | Group | Label | Coords | Faces | Type |
|---|---|---|---|---|---|
| | 1 | A | -7, -4, 2 | AFKG | 1 |
| | 1 | B | -7, 4, -2 | AIEB | 2 |
| | 1 | C | 7, -2, -4 | BELH | 1 |
| | 1 | D | 7, 2, 4 | CDHL | 1 |
| | 2 | E | -1, -2, -8 | DCGK | 1 |
| | 2 | F | -1, 2, 8 | CEIG | 2 |
| | 2 | G | 1, -8, 2 | ABJF | 2 |
| | 2 | H | 1, 8, -2 | DHJF | 2 |
| | 3 | I | -3, -6, -3 | AGI | 3 |
| | 3 | J | -3, 6, 3 | BHJ | 3 |
| | 3 | K | 3, -3, 6 | CLE | 3 |
| | 3 | L | 3, 3, -6 | DKF | 3 |

*Fig. 3: integer coordinates*

The vertices in groups 1 and 2 have valence 4 (3 quads and one of the acute angles of the triangles) and are on the circumsphere at $R_2$, the vertices in Group 3 have valence 3 (1 obtuse angle of the triangles and 2 obtuse angles of the quads) and are at $R_1$ from the center. The *Josehedron* has two different inspheres, one touching the four triangles at $D_1=\sqrt{31.5}\cong 5.612$, and one touching the eight quads at $D_2=\sqrt{45}\cong 6.708$ (see Fig. 4). The 12 faces consist of 4 isosceles triangles (face type 3) and 8 quadrilaterals, which come in two mirror-symmetric variants: 4 left-handed (type 1) and 4 right-handed (type 2).

*Table 2: closest points on faces from origin, also face normal vectors, point shared with neighbor shape through Delaunay edge, point where insphere touches, column Ri is the insphere radius index from Fig. 4*

| X | Y | Z | $R_i$ |
|---|---|---|---|
| -4.5 | -1.5 | -3 | 4 |
| -4.5 | 1.5 | 3 | 4 |
| -3 | -6 | 0 | 3 |
| -3 | 6 | 0 | 3 |
| -1.5 | -3 | 4.5 | 4 |
| -1.5 | 3 | -4.5 | 4 |
| 1.5 | -4.5 | -3.0 | 4 |
| 1.5 | 4.5 | 3 | 4 |
| 3 | 0 | 6 | 3 |
| 3 | 0 | -6 | 3 |
| 4.5 | -3 | 1.5 | 4 |
| 4.5 | 3 | -1.5 | 4 |

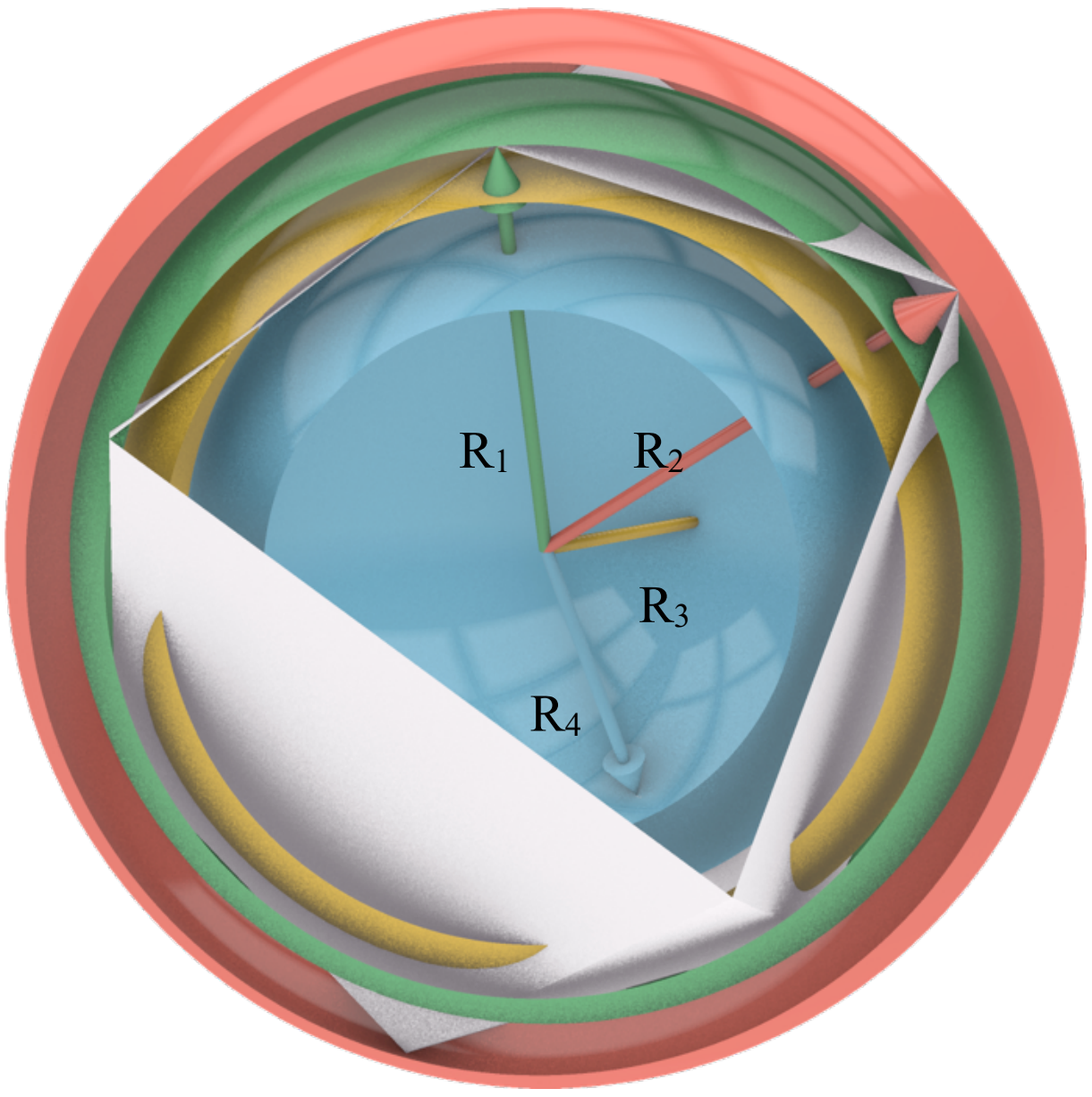

*Fig. 4: two circumspheres (red $R_2$, green $R_1$) and two inspheres (yellow $R_3$, blue $R_4$)*

The regular dodecahedron (equal number of faces as the *Josehedron*) fills ~66.49% of the volume of its circumsphere (the sphere through all vertices), and a regular icosahedron, the maximum-volume polyhedron with 12 vertices (equal number of vertices as the *Josehedron*) inscribed in a sphere, occupies ~60.55% thereof. The *Josehedron* fills ~47.98% of the volume of its outer circumsphere. This is slightly superior to another SFPH, the rhombic dodecahedron at ~47.75%, though the latter has 16 vertices. The cube, in comparison, only occupies ~36.76% of its circumsphere. The maximal hexagonal prism (also an SFPH, and a suspected candidate by reddit user "st3f-ping"[28]) inscribed in a unit sphere occupies ~47.75%. The *Josehedron* is therefore, by a small margin, the "roundest" SFPH known to date.

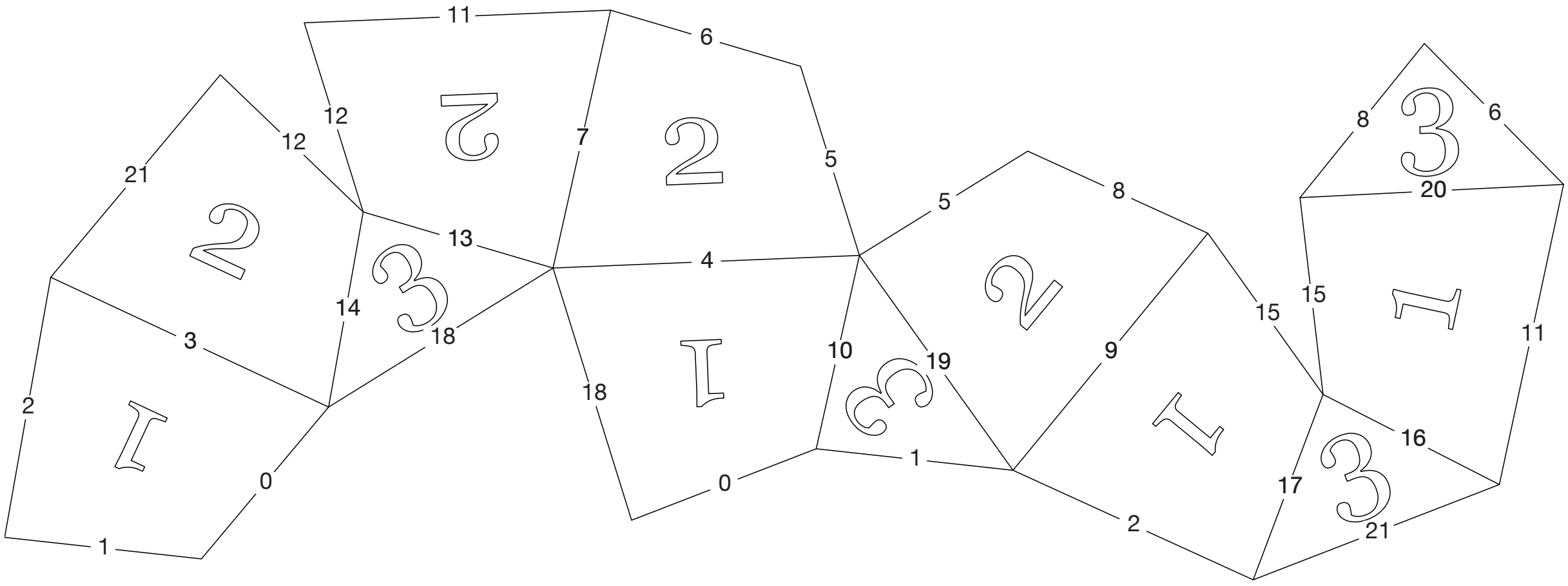

*Fig. 5: unrolled version of the Josehedron, big numbers indicate face type and orientation, small numbers show corresponding edge index numbers (0-21)*

The 22 edges of the *Josehedron* come in three distinct lengths (see Table 3). Two of the interior angles of each quadrilateral face (types 1 and 2), specifically those adjacent to the longest edge (3,4,9,11), are equal (~75.037°). The four faces of type 1 form a continuous developable strip, joined by edges 0, 2, and 15, so do the faces of type 2, joined by edges 5, 7, and 12. These two spatial U-shapes are combined, like the two parts of a tennis ball, with the resulting gaps filled by the four triangles (type 3). This is shown by the face colors in Fig. 1. The faces of type 1 are surrounded, in counterclockwise order, by edges BCAA, type 2 by edges BAAC, and type 3 by edges ACA.

*Table 3: edge properties*

| Label | Length | Count | Edge indices (Fig. 5) |
|---|---|---|---|
| A | ~ 6.708 | 12 | 0, 1, 5, 6, 8, 10, 12, 13, 14, 15, 16, 17 |
| B | ~ 10.392 | 4 | 3, 4, 9, 11 |
| C | ~ 8.944 | 6 | 2, 7, 18, 19, 20, 21 |

The *Josehedron* described above was generated from the minima of the FKS function. Using the maxima instead produces the same shape, but as a chiral (mirror-image) copy, obtainable by reflecting the polyhedron shown in Fig. 3 about any of the world planes XY, YZ, or ZX. When the minima and maxima are combined into a single point set and the Voronoi tessellation is recomputed, yet another space-filling polyhedron emerges. This combined plesiohedron has 14 faces and 16 vertices (Fig. 6), is more irregular than the *Josehedron*, and one needs the two chiral types (mirror copies of each other) to fill space.

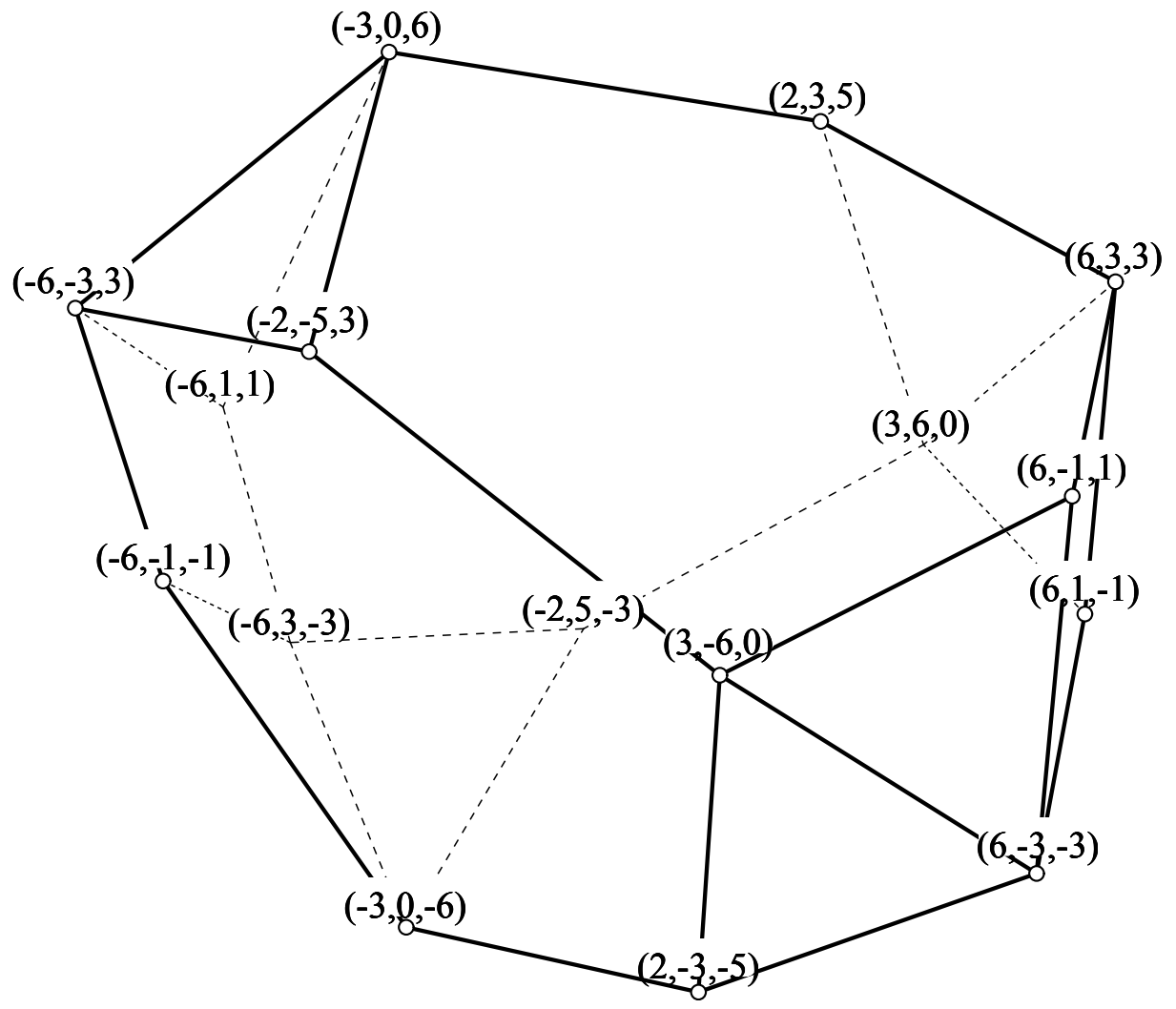

*Fig. 6: space-filling polyhedron of the point set containing both minima and maxima of the FKS function*

## 3.2. Center Point Distribution

The coordinate values of the 12 extremal points listed in Table 4, the minima of the function (Eq. 1), in the unit cell $[0,1\}^3$ are all fractions, multiples of 1/8. The following table shows the numerators (example: 3 = 3/8 = 0.375). To obtain the minimum and maximum values ±√2 of the function, the values V for X, Y, and Z must be transformed by $V/8 * 2\pi$. To obtain the integer coordinates for the resulting plesiohedra's vertices shown in Fig. 3, the shape needs to be scaled by a factor of 3.

*Table 4: point coordinates for minima and maxima, column i refers to the list in chapter "unit cell" (colors and rotations)*

| # | X | Y | Z | i | X | Y | Z |
|---|---|---|---|---|---|---|---|
| 1 | 0 | 2 | 3 | 6 | 0 | 2 | 7 |
| 2 | 0 | 6 | 1 | 5 | 0 | 6 | 5 |
| 3 | 1 | 0 | 6 | 1 | 1 | 4 | 2 |
| 4 | 2 | 3 | 0 | 3 | 2 | 1 | 4 |
| 5 | 2 | 5 | 4 | 4 | 2 | 7 | 0 |
| 6 | 3 | 0 | 2 | 2 | 3 | 4 | 6 |
| 7 | 4 | 2 | 5 | 5 | 4 | 2 | 1 |
| 8 | 4 | 6 | 7 | 6 | 4 | 6 | 3 |
| 9 | 5 | 4 | 2 | 1 | 5 | 0 | 6 |
| 10 | 6 | 1 | 0 | 4 | 6 | 3 | 4 |
| 11 | 6 | 7 | 4 | 3 | 6 | 5 | 0 |
| 12 | 7 | 4 | 6 | 2 | 7 | 0 | 2 |
| | | minima | | | | maxima | |

The extremal points, the centers each *Josehedron* is formed around, form a Delone set[29]. The packing radius *r* of this set is √3.5≅1.871, or half the distance between any point and its closest neighbor. The covering radius *R* of this set, the smallest distance such that every point in space is within distance *R* of at least one point, is 2.769 (√69/3), the radius of the outer circumsphere.

When the 12 extremal points are projected onto any of the three coordinate planes (XY, YZ, or XZ), they coincide exactly with the nodes of a pentagonal Cairo tiling (Fig. 7, left). The pentagons of this Cairo tiling have remarkably regular proportions (Fig. 7, right): the base-to-width ratio is 1:2, the distance from the base AB to the two right-angle corners C and E is also 1, and the four remaining sides are of equal length $a$, which is

$$a = \sqrt[2]{1^2 + \left({}^{1}/_{2}\right)^2} \approx 1.118$$

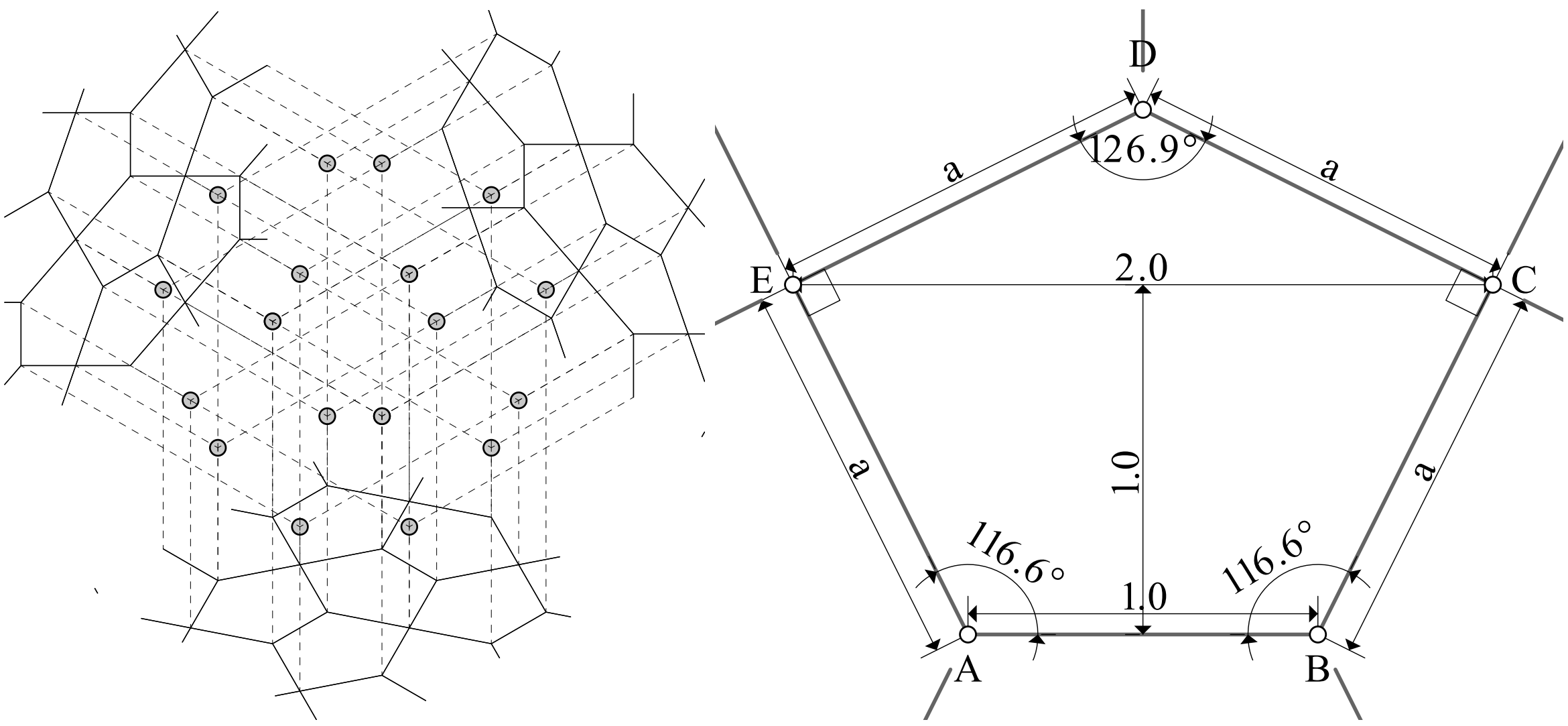

*Fig. 7: left: points projected on world planes lie on the nodes of a Cairo tiling; right: Cairo pentagonal tile dimensions*

The individual shapes' centers forming the nodes of a Cairo tiling make the *Josehedron* in a way the dual of the bisymmetric hendecahedron[30,31], as well as Williams' β-Tetrakaidecahedron[32], whose projected edges form one hexagon made of four pentagons of a Cairo tiling.

When projected along any of the diagonals (±1,±1,±1) on a plane normal to the latter, the points also exhibit another interesting pattern (Fig. 8). Notice that three of the blue polygons form equilateral triangles, with one node in the triangle's center, but the other nodes do not bisect the triangles' edges. This is because the red polygons do not form equilateral hexagons. It is possible that if the points along the corresponding diagonals were moved to the corners of equilateral hexagons, they would also form a Delone set and the corresponding plesiohedra would also be a novel type of SFPH.

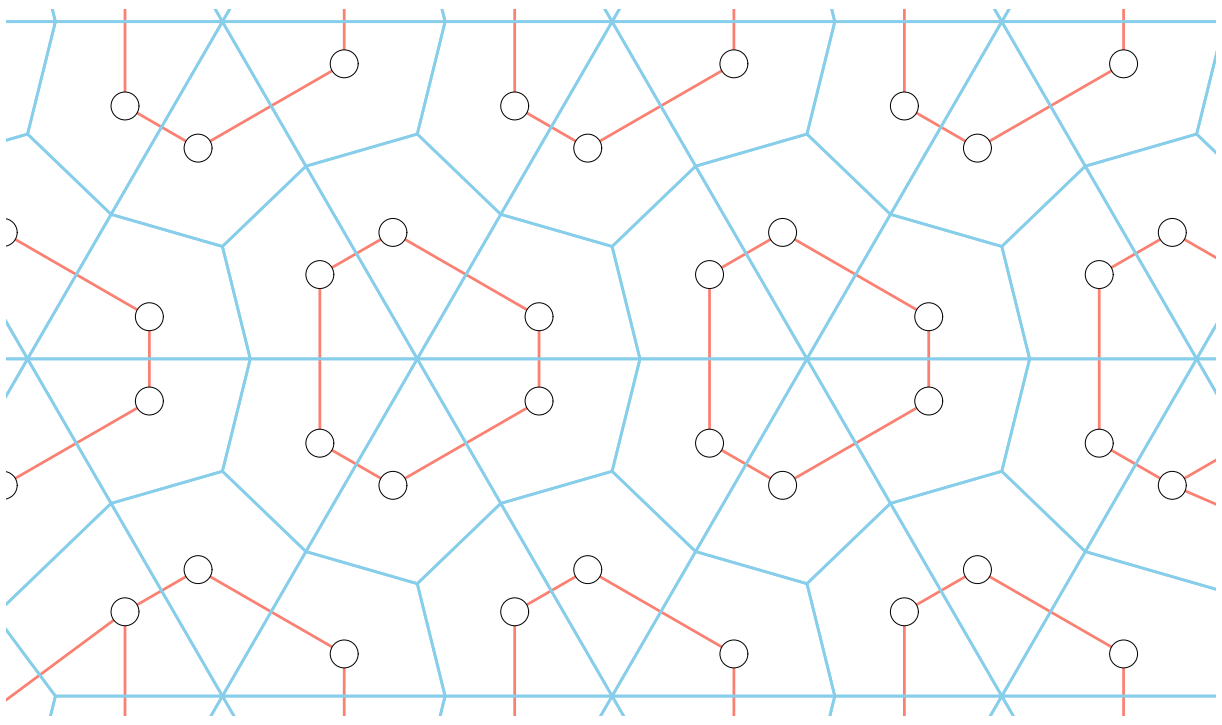

*Fig. 8: points project along a spatial diagonal (black), Voronoi pattern (blue), connection to 2 closest neighbors (red)*

For the Gyroid TPMS, progressive thinning (by adding a constant value or lowering the iso-surface threshold) eventually yields a tubular structure that is topologically equivalent to the Laves graph. However, this elegant correspondence does not hold for the FKS surface. As illustrated in the figure below, the skeleton of the FKS has a different connectivity, with more branching points than the Laves graph.

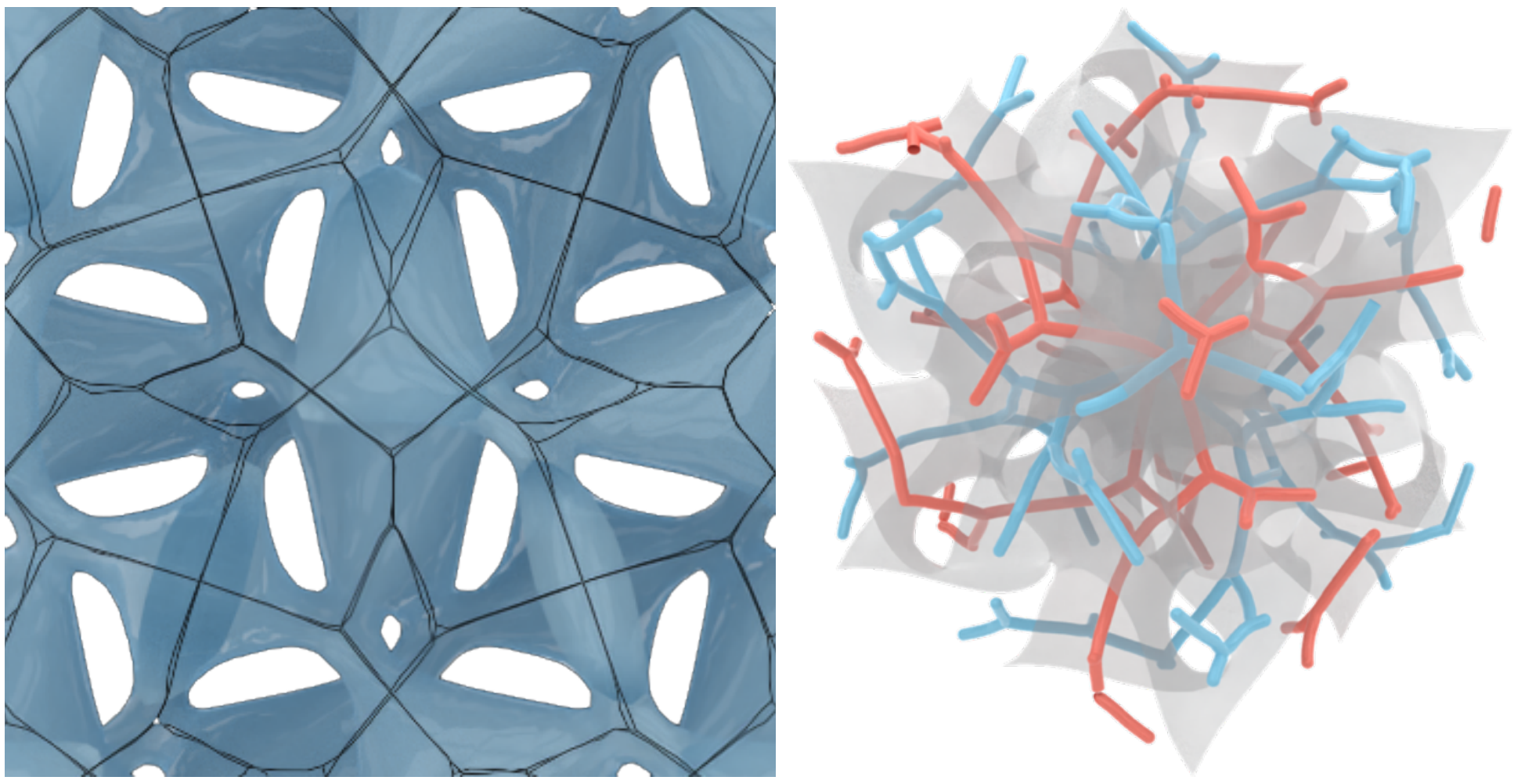

*Fig. 9: left: FKS thinned by 0.6 (blue) and its medial axis skeleton lines (black); right: interwoven skeleton graphs of the FKS TPMS*

The two skeletons, understood as the medial axes on either side of the FKS TPMS, are mirror-symmetric, interwoven, and never touching, and feature nodes of valence 3 connected by straight edges, in the same way that the Laves graph is the skeleton of the Gyroid TPMS. The skeleton of the Diamond TPMS has nodes of valence 4 (tetrahedral), forming a diamond cubic lattice of carbon atoms.

## 3.3. Unit cell

The cubic unit cell contains 12 instances of the *Josehedron*, arranged in 6 distinct orientations with 2 copies of each orientation. This conglomerate block of 12 shapes has translational symmetry along all three spatial directions X, Y, and Z.

| Top (-Z) | Front (Y) | Left (X) |
|---|---|---|
| 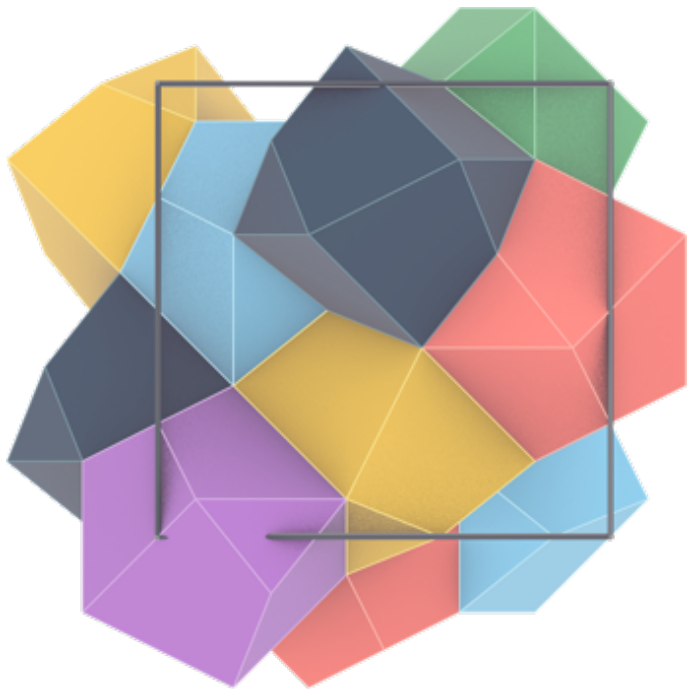 | 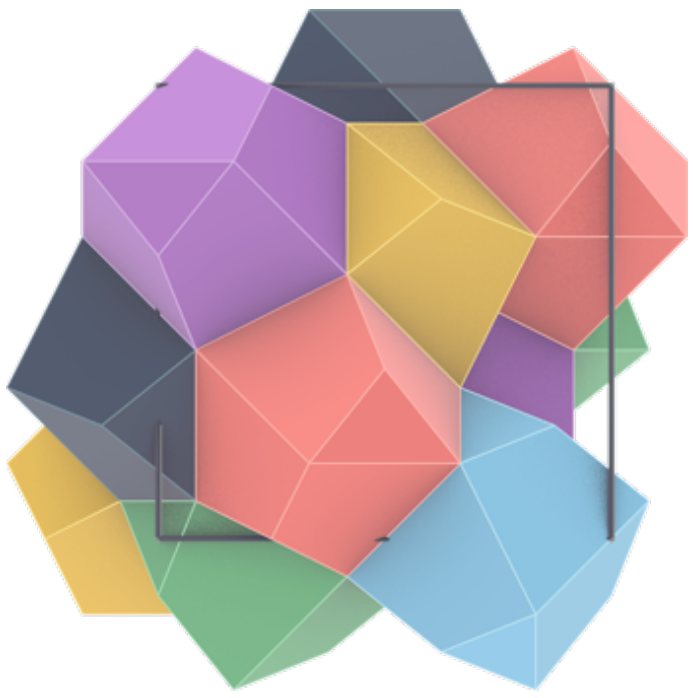 | 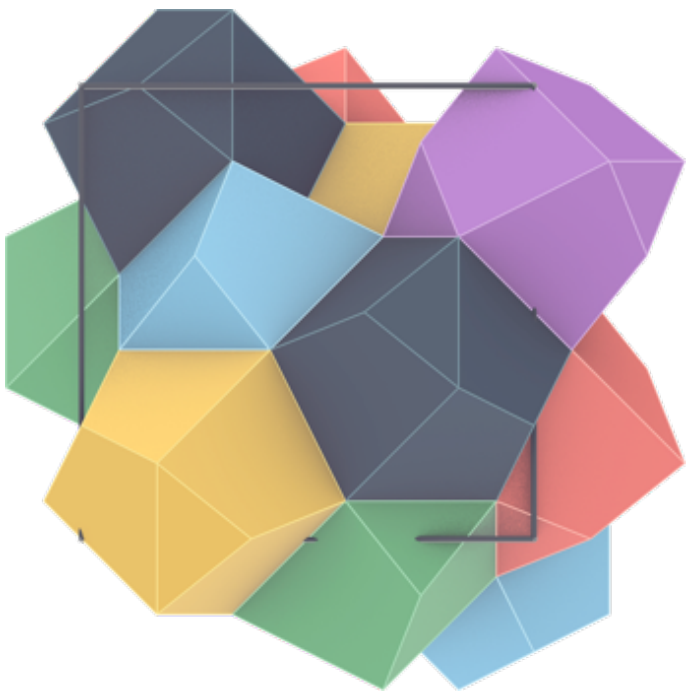 |
| **Bottom (Z)** | **Back (-Y)** | **Right (-X)** |
| 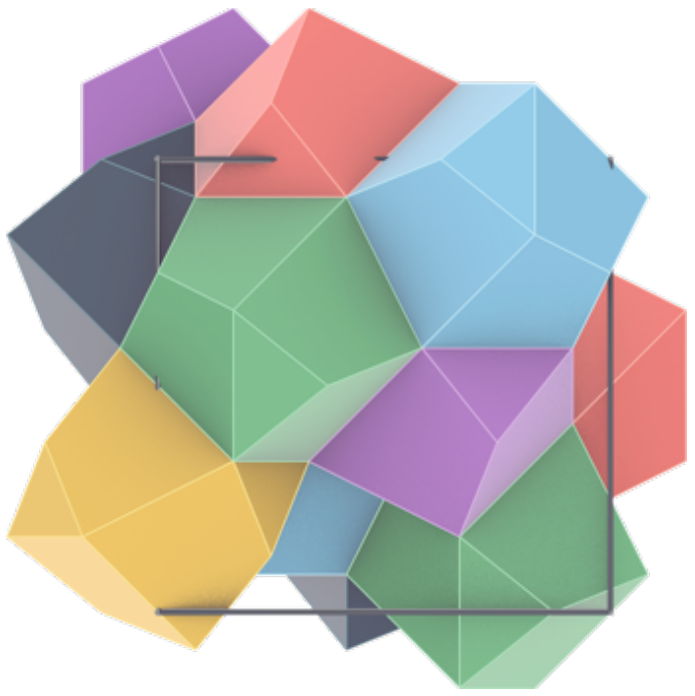 | 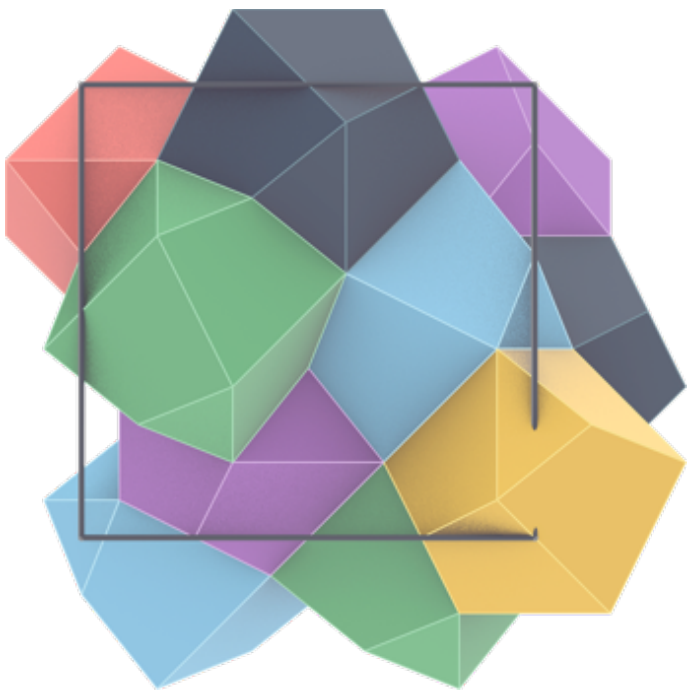 | 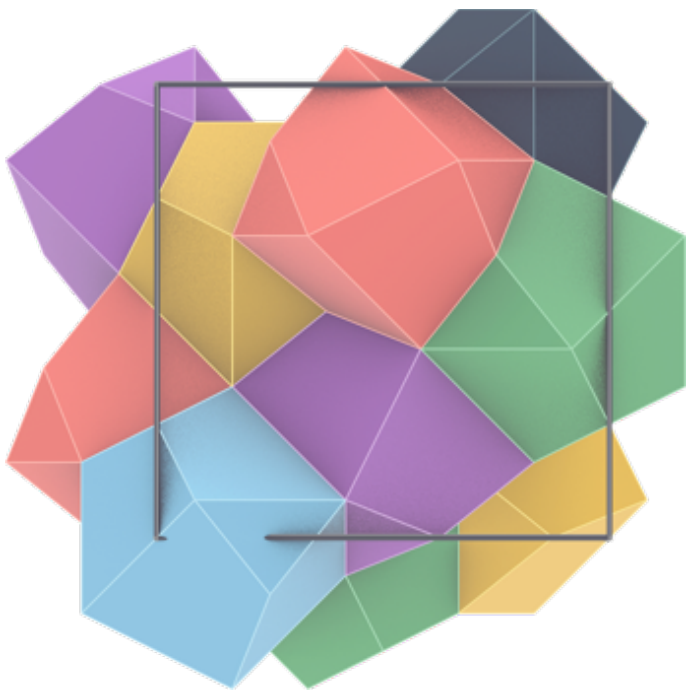 |

*Fig. 10: orthographic and 6 side views of the unit cell, view direction in parentheses*

The centers of the 12 shapes are located at the integer coordinate positions in Table 4, their orientations are achieved through a sequence of rotations (π/2 or π) around the world axes, and not the object coordinate system. The rotations for each color are shown in Table 5. It is possible that some of the above rotations can be reached with other – simpler – axis combinations, but these proved to work empirically.

*Table 5: Rotations of the individual shapes in the unit cell*

| # | Color | Rotations |
|---|---|---|
| 1 | Lavender | unmodified, as described in Shape |
| 2 | Red | 180° Z |
| 3 | Green | -90° Y → 90° X → 180° Z |
| 4 | Sky Blue | -90° Y → 90° X |
| 5 | Yellow | 90° X → 90° Y |
| 6 | Dark Blue | 90° X → -90° Y |

A larger assembly of Josehedra into a truly space-filling arrangement of 144 unit cells (12 per unit cell, 4 translated copies in X, 3 translated copies in Y) is shown in Fig. 11.

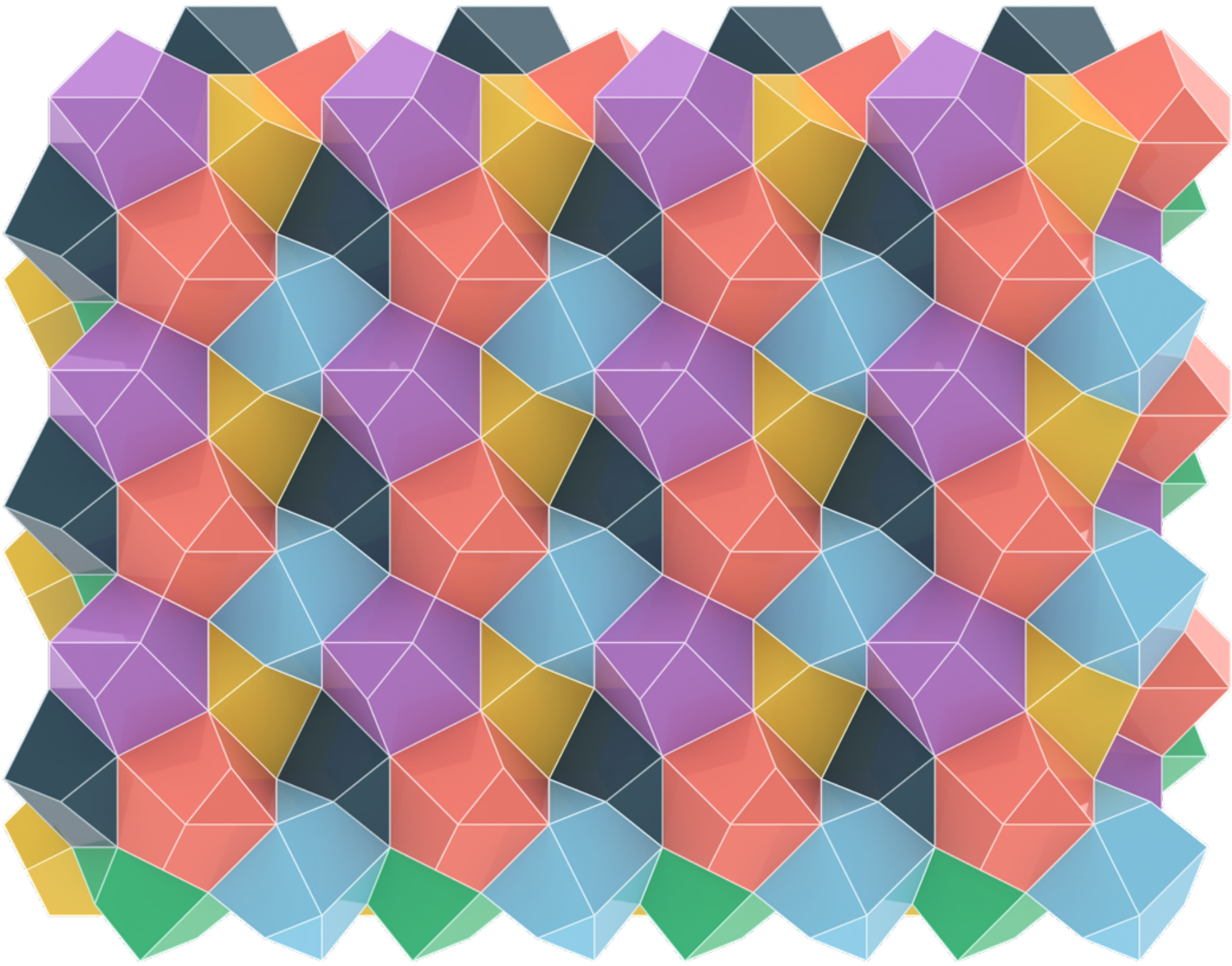

*Fig. 11: assembly 4 by 3 unit cells, front view*

# 4. Discussion

The method presented here can be extended to many other TPMS beyond the FKS surface. Dozens of TPMS are now catalogued in literature and collected e.g. by Rhino user corax, who lists approximately 40 surfaces[33,34]. Given the relative simplicity of the construction and the elegance of the resulting polyhedron, it would be surprising if the *Josehedron* had never been encountered before. However, to the best of the author's knowledge, no prior documentation or depiction of this specific plesiohedron exists in the literature.

Furthermore, the family of triply periodic surfaces is not limited to the classically catalogued examples. As long as certain symmetry constraints are respected – namely, that the spatial coordinates appear as integer multiples, that all three spatial directions are treated equivalently, and that only scalar multipliers are applied – any combination of sine and cosine functions through addition, subtraction, multiplication, or division can in principle define a valid triply periodic surface whose extremal points may yield further novel plesiohedra, cousins in the family of the *Josehedron*. A basis for such an exploration is provided in the appendix, in the chapter Blocks. These can be used to put together new triply periodic functions (not guaranteed to result in minimal surfaces, but for our method of constructing Voronoi cells around extrema points, also not necessary) by trial and error, brute-force combinatorics, or the help of a creative LLM-based agent.

# 5. Acknowledgement

I would like to thank the Digital Building Technologies (DBT) group for providing an amazing work environment and all my fantastic DBT colleagues for helping me with 3D-printing prototypes of the *Josehedon*.

# 6. References

1. Ruzicka, E. O. Morphology of polyhedral space habitat modules – Identifying the ideal form using multi-criteria analysis. *Acta Astronautica* **221**, 66–78 (2024).
2. Sabariego, P. & Santos, F. On the number of facets of three-dimensional Dirichlet stereohedra IV: quarter cubic groups. *Beitr Algebra Geom* **52**, 237–263 (2011).
3. Yokoyama, T., Ichikawa, K. & Naito, H. From polyhedra to crystals: a graph-theoretic framework for crystal structure generation. *CrystEngComm* 10.1039.D5CE01176K (2026) doi:10.1039/D5CE01176K.
4. Lagarias, J. C. & Zong, C. Mysteries in Packing Regular Tetrahedra. *Notices Amer. Math. Soc.* **59**, 1392 (2012).
5. Weisstein, E. W. Space-Filling Polyhedron. https://mathworld.wolfram.com/Space-FillingPolyhedron.html.
6. Space-filling polyhedron. *Wikipedia* (2026).
7. Wells, D. G. *The Penguin Dictionary of Curious and Interesting Geometry*. (Penguin Books, London, 1991).
8. Raumfueller - regulaere Raumfuellung. http://www.3doro.de/space-filling/.
9. Gómez-Gálvez, P. *et al.* Scutoids are a geometrical solution to three-dimensional packing of epithelia. *Nat Commun* **9**, 2960 (2018).
10. Weaire, D. & Phelan, R. A counter-example to Kelvin's conjecture on minimal surfaces. *Philosophical Magazine Letters* **69**, 107–110 (1994).
11. Goldberg, M. On the dodecahedral space-fillers. *Geom Dedicata* **10**, 79–89 (1981).
12. Engel, P. Über Wirkungsbereichsteilungen von kubischer Symmetrie. *Zeitschrift für Kristallographie - Crystalline Materials* **157**, 259–276 (1981).
13. Doroziński, T. E. Plesiohedron. *Geometryka* https://geometryka.wordpress.com/2021/03/08/plesiohedron/ (2021).
14. Grünbaum, B. & Shephard, G. C. Tilings with congruent tiles. *Bull. Amer. Math. Soc.* **3**, 951–973 (1980).
15. Grünbaum, B. & Shephard, G. C. 69.14 Space filling with identical symmetrical solids. *Math. Gaz.* **69**, 117–120 (1985).
16. Fisher, J. jwf23/Equation-Based-Lattice-Structure-Dataset. (2026).
17. Vasile, A., Constantinescu, D. M., Coropețchi, I. C., Sorohan, Ștefan & Apostol, D. A. Definition, Fabrication, and Compression Testing of Sandwich Structures with Novel TPMS-Based Cores. *Materials* **17**, 5150 (2024).
18. Shi, J. *et al.* A TPMS-based method for modeling porous scaffolds for bionic bone tissue engineering. *Sci Rep* **8**, 7395 (2018).
19. Hsieh, M.-T. & Valdevit, L. Minisurf – A minimal surface generator for finite element modeling and additive manufacturing. *Software Impacts* **6**, 100026 (2020).

20. Chris-Amadin, H. & Ibhadode, O. LattGen: A TPMS lattice generation tool. *Software Impacts* **21**, 100665 (2024).
21. Fischer, W. & Koch, E. On 3-periodic minimal surfaces. *Zeitschrift für Kristallographie - Crystalline Materials* **179**, 31–52 (1987).
22. Akbari, M., Mirabolghasemi, A., Akbarzadeh, H. & Akbarzadeh, M. Geometry-based structural form-finding to design architected cellular solids. in *Symposium on Computational Fabrication* 1–11 (ACM, Virtual Event USA, 2020). doi:10.1145/3424630.3425419.
23. Akbari, M., Lu, Y. & Akbarzadeh, M. From Design to the Fabrication of Shellular Funicular Structures. in 328–339 (Online and Global, 2021). doi:10.52842/conf.acadia.2021.328.
24. Akbari, M., Mirabolghasemi, A., Bolhassani, M., Akbarzadeh, A. & Akbarzadeh, M. Strut-Based Cellular to Shellular Funicular Materials. *Adv Funct Materials* **32**, 2109725 (2022).
25. Akbarzadeh, M. *et al.* Diamanti: 3D-printed, post-tensioned concrete canopy. in *Creating Resourceful Futures* (Copenhagen, Denmark, 2024). doi:https://doi.org/10.2307/jj.11374766.40.
26. Chai, H., Bernhard, M., Zhi, Y., Ororbia, M. E. & Akbarzadeh, M. Design-to-fabrication workflow for 3D concrete printed structures with embedded periodic anticlastic surfaces. *Automation in Construction* **185**, 106855 (2026).
27. Bernhard, M., Dillenburger, B. & Clémente, R. Axolotl. Digital Building Technologies / ETH Zurich (2018).
28. Tuskadaemonkilla. Which space-filling polyhedron has the smallest surface area to volume ratio? *r/mathematics* https://www.reddit.com/r/mathematics/comments/z6slah/which_spacefilling_polyhedron_has_the_smallest/ (2022).
29. Delone set. *Wikipedia* (2026).
30. Inchbald, G. Five Space-Filling Polyhedra. https://www.steelpillow.com/polyhedra/five_sf/five.html.
31. Wu, J. & Inchbald, G. Folding Space-Filling Bisymmetric Hendecahedron for a Large-Scale Art Installation. in *Proceedings of Bridges 2018: Mathematics, Art, Music, Architecture, Education, Culture* 483–486 (2018).
32. Inchbald, G. Williams β-Tetrakaidecahedron. https://www.steelpillow.com/polyhedra/Williams/williams.html.
33. corax. Isopod - Extended TPMS component. *McNeel Forum* https://discourse.mcneel.com/t/isopod-implicit-surface-tools/188951/258 (2025).
34. corax. Isopod - TPMS collection. *McNeel Forum* https://discourse.mcneel.com/t/isopod-implicit-surface-tools/188951/373 (2026).
35. Walles, J. What equations are used to create the TPMS types? *nTop Support* https://support.ntop.com/hc/en-us/articles/360053267814-What-equations-are-used-to-create-the-TPMS-types (2025).

# 7. Appendix

## 7.1. TPMS Formulae

*Table 6: The formulae for TPMS as used by Axolotl and in this study*

| Name | Formula |
|---|---|
| Diamond | $\sin x \sin y \sin z + \sin x \sin y \cos z + \cos x \sin y \cos z + \cos x \cos y \sin z$ |
| Double Diamond | $\sin 2x \sin 2y + \sin 2y \sin 2z + \sin 2z \sin 2x + \cos 2x \cos 2y \cos 2z$ |
| Double Gyroid | $2.75(\sin 2x \sin z \cos y + \sin 2y \sin x \cos z + \sin 2z \sin y \cos x) - (\cos 2x \cos 2y + \cos 2y \cos 2z + \cos 2z \cos 2x)$ |
| DP | $0.5(\cos x \cos y + \cos y \cos z + \cos z \cos x) + 0.2(\cos 2x \cos 2y \cos 2z)$ |
| Fischer-Koch, FKS | $\cos 2x \sin y \cos z + \cos 2y \sin z \cos x + \cos 2z \sin x \cos y$ |
| FRP | $8 \cos x \cos y \cos z + \cos 2x \cos 2y \cos 2z - (\cos 2x \sin 2y + \cos 2y \sin 2z + \cos 2z \sin 2x)$ |
| Gyroid | $\sin x \cos y + \sin y \cos z + \sin z \cos x$ |
| IWP | $2(\cos x \cos y + \cos y \cos z + \cos z \cos x) - (\cos x^2 + \cos y^2 + \cos z^2)$ |
| KP | $0.6(\cos x + \cos y + \cos z) + 0.7(\cos x \cos y + \cos y \cos z + \cos z \cos x) - 0.9(\cos 2x \cos 2y \cos 2z) + 0.4(\cos x + \cos y + \cos z) + 0.7(\cos x \cos y + \cos y \cos z + \cos z \cos x) - 0.9(\cos 2x \cos 2y \cos 2z) + 0.4$ |
| Lidinoid | $0.5(\sin 2x \cos y \sin z + \sin 2y \cos z \sin x + \sin 2z \cos x \sin y) - 0.5(\cos 2x \cos 2y + \cos 2y \cos 2z + \cos 2z \cos 2x) + 0.15$ |
| Neovius | $3(\cos x + \cos y + \cos z) + 4 \cos x \cos y \cos z$ |
| Octo | $4(\cos x \cos y + \cos y \cos z + \cos z \cos x) - 2.8(\cos x \cos y \cos z) + \cos x + \cos y + \cos z + 1.5$ |
| Schwarz P | $\cos x + \cos y + \cos z$ |
| Split P[35] | $1.1(\sin 2x \sin z \cos y + \sin 2y \sin x \cos z + \sin 2z \sin y \cos x) - 0.2(\cos 2x \cos 2y + \cos 2y \cos 2z + \cos 2z \cos 2x) - 0.4(\cos 2x + \cos 2y + \cos 2z)$ |

The formulae currently implemented in Axolotl (see Table 6) have been collected over several years from multiple sources. Their naming and exact formulation may vary between different authors and provenances. Rhino and Grasshopper user "corax" has put together an even more expansive list of 45 different TPMS formulae (see Table 7), which served as the basis for the following analysis of blocks.

*Table 7: TPMS formulae collected by user "corax"*

| Name | Alternative Names | f(x, y, z) |
|---|---|---|
| **Standard Surfaces** | | |
| Gyroid | G, Schoen G, Schoen Gyroid, Y*, Y** | cos x sin y + cos y sin z + cos z sin x |
| D Surface | Schwarz D, Diamond D, D*, Black D | sin x sin y sin z + sin x cos y cos z + cos x sin y cos z + cos x cos y sin z<br>x,y,z at half frequency |
| Diamond | — | cos x cos y cos z − sin x sin y sin z<br>x,y,z at half frequency |
| P Surface | Primitive, Schwarz P, Simple Cubic, P* | cos x + cos y + cos z |
| C(D) Surface | C(D*), Complementary D | cos (3x+y) cos z − sin (3x−y) sin z + cos (x+3y) cos z + sin (x−3y) sin z + cos (x−y) cos 3z − sin (x+y) sin 3z<br>x,y,z at half frequency |
| IWP | I-WP, Surface BCC, IP2-J*, Schoen I-graph-wrapped package | 2(cos x cos y + cos y cos z + cos z cos x) − (cos 2x + cos 2y + cos 2z) |
| Fisher-Koch | — | (cos x cos y + cos y cos z + cos z cos x) − (cos 2x + cos 2y + cos 2z) |
| Lidinoid | HG, L | sin 2x sin z cos y + sin 2y sin x cos z + sin 2z sin y cos x − (cos 2x cos 2y + cos 2y cos 2z + cos 2z cos 2x) + 0.3<br>x,y,z at half frequency |
| OCTO | O,C-TO, O,CT-O | 0.6(cos x cos y + cos y cos z + cos z cos x) − 0.4(cos x + cos y + cos z) + 0.25 |
| FRD | F-RD, Fxx-P2Fz, Schoen FRD | $8\cos x\cos y\cos z + \cos 2x\cos 2y\cos 2z - (\cos 2x\cos 2y + \cos 2x\cos 2z + \cos 2y\cos 2z)$ |
| FRD Prime | S6-Surface | $4\cos x\cos y\cos z - (\cos 2x\cos 2y + \cos 2x\cos 2z + \cos 2y\cos 2z)$ |
| S Surface | Fischer Koch S, S* | $\cos 2x\sin y\cos z + \cos 2y\sin z\cos x + \cos 2z\sin x\cos y$ |
| C(S) Surface | Complementary S, Fisher-Koch C(S) | $\cos 2x + \cos 2y + \cos 2z + 2(\sin 3x\sin 2y\cos z + \cos x\sin 3y\sin 2z + \sin 2x\cos y\sin 3z) + 2(\sin 2x\cos 3y\sin z + \sin x\sin 2y\cos 3z + \cos 3x\sin y\sin 2z)$ |
| PN Surface | P+C(P) | $0.3\cos x\cos y\cos z + 0.2(\cos x + \cos y + \cos z) + 0.1\cos 2x\cos 2y\cos 2z + 0.1(\cos 2x + \cos 2y + \cos 2z) + 0.05\cos 3x\cos 3y\cos 3z + 0.1(\cos x\cos y + \cos y\cos z + \cos z\cos x)$ |
| G Prime 1 | G', G Prime 2 [Chr2024] | $\sin 2x\sin z\cos y + \sin 2y\sin x\cos z + \sin 2z\sin y\cos x + \cos 2x\cos 2y + \cos 2y\cos 2z + \cos 2z\cos 2x + 0.32$ |
| G Prime 2 | $G'_2$, G Prime 1 [Chr2024] | $5(\sin 2x\sin z\cos y + \sin 2y\sin x\cos z + \sin 2z\sin y\cos x) + \cos 2x\cos 2y + \cos 2y\cos 2z + \cos 2z\cos 2x$ |
| D Prime | D' | $0.5(\cos x\cos y\cos z + \cos x\sin y\sin z + \sin x\cos y\sin z + \sin x\sin y\sin z) - 0.5(\sin 2x\sin 2y + \sin 2y\sin 2z + \sin 2z\sin 2x) - 0.2$ |
| K Surface | Karcher K, KP | $0.3(\cos x + \cos y + \cos z) + 0.3(\cos x\cos y + \cos y\cos z + \cos z\cos x) - 0.4(\cos 2x + \cos 2y + \cos 2z) + 0.2$ |
| Y Surface | Fisher-Koch Y | $\cos x\cos y\cos z + \sin x\sin y\sin z + \sin 2x\sin y + \sin 2y\sin z + \sin 2z\sin x + \cos x\sin 2y + \cos y\sin 2z + \cos z\sin 2x$ |
| C(Y) Surface | Fisher-Koch C(Y), Complementary Y, (YYxxx)* | $\cos x\cos y\cos z - \sin x\sin y\sin z + \sin 2x\sin y + \sin 2y\sin z + \sin x\sin 2z - \sin 2x\cos z + \sin 2y\cos x + \sin 2z\cos y$ |
| PMY | +/−Y, (Fxxx)* | $2\cos x\cos y\cos z + \sin 2x\sin y + \sin 2y\sin z + \sin 2z\sin x$ |
| CPMY | Complementary +/−Y, (FFxxx)* | $-2\cos x\cos y\cos z + \sin 2x\sin y + \sin 2y\sin z + \sin 2z\sin x$ |
| C(I2-Y**) | Complementary I2-Y**, S*-Yxxx** | $2(\sin 2x\cos y\sin z + \sin x\sin 2y\cos z + \cos x\sin y\sin 2z) + \cos 2x\cos 2y + \cos 2y\cos 2z + \cos 2x\cos 2z$ |
| Q* Surface | — | $(\cos x - 2\cos y)\cos z - \sqrt{3}\sin z\,(\cos(x-y) - \cos x) + \cos(x-y)\cos z$ |
| C(G) Surface | Complementary G, C(Y**), Complementary Y** | $3(\sin x\cos y + \sin y\cos z + \sin z\cos x) + 2(\sin 3x\cos y + \sin 3y\cos z + \sin 3z\cos x) - 2(\sin x\cos 3y + \sin y\cos 3z + \sin z\cos 3x)$ |
| Bionic Bone 1 | — | $20(\cos x\sin y + \cos y\sin z + \cos z\sin x) - 0.5(\cos 2x\cos 2y + \cos 2y\cos 2z + \cos 2z\cos 2x) - 4$ |
| Bionic Bone 2 | — | $10(\cos x\sin y + \cos y\sin z + \cos z\sin x) - 2(\cos 2x\cos 2y + \cos 2y\cos 2z + \cos 2z\cos 2x) - 12$ |
| Neovius | C9(P), C(P), Complementary P, P*, J* | $3(\cos x + \cos y + \cos z) + 4\cos x\cos y\cos z$ |
| Slotted-P | WzI-Wxx | $-2(\cos x\cos y + \cos y\cos z + \cos z\cos x) - 2(\cos 2x + \cos 2y + \cos 2z) + (\cos 2x\cos y + \cos 2y\cos z + \cos 2z\cos x) - (\cos x\cos 2y + \cos y\cos 2z + \cos z\cos 2x)$ |
| S3 Surface | — | $\cos x\cos y + \cos x\cos z + \cos y\cos z + \sin x\cos y + \sin x\cos z + \sin y\cos z + \sin y\cos z + \sin z\cos x + \sin z\cos y$ |
| S4 Surface | — | $\cos 2x\cos y\cos z + \cos 2y\cos x\cos z + \cos 2z\cos x\cos y + \sin x\cos y + \sin x\cos z + \sin y\cos z + \sin y\cos x + \sin z\cos x + \sin z\cos y$ |
| S7 Surface | — | $4\sin x\cos y\cos z - (\cos x\cos y + \cos y\cos z + \cos z\cos x)$ |
| S8 Surface | — | $8(\cos x\cos z\sin x + \cos y\cos z\sin y + \cos x\cos y\sin z)$ |
| S9 Surface | — | $3\sin x\sin y - 4\cos x\cos y\cos z$ code has 3× sin x sin y (likely a bug, taken directly from Vasile et al. [17], Table 1, Equation 9. should probably be pairwise cyclic: xy+yz+zx) |

| Double Surfaces | | |
|---|---|---|
| Double Gyroid | — | $2.75(\sin 2x \cos y \sin z + \sin 2y \cos z \sin x + \sin 2z \cos x \sin y) - (\cos 2x \cos 2y + \cos 2y \cos 2z + \cos 2z \cos 2x)$ normalizer only on 2nd term in code (possible bracket bug) |
| I2-Y** | — | $-2(\sin 2x \cos y \sin z + \sin x \sin 2y \cos z + \cos x \sin y \sin 2z) + \cos 2x \cos 2y + \cos 2y \cos 2z + \cos 2x \cos 2z$ |
| Double D | Double Diamond, Double Diamond 1 | $\sin x \sin y + \sin y \sin z + \sin z \sin x + \cos x \cos y \cos z$ |
| Double D 2 | Double Diamond 2 | $\cos x \cos y + \cos y \cos z + \cos z \cos x + \sin x \sin y \sin z$ |
| Double P | DP-Surface, Double Primitive | $0.5(\cos x \cos y + \cos y \cos z + \cos z \cos x) + 0.2(\cos 2x + \cos 2y + \cos 2z)$ |
| Split P | P2-DG | $1.1(\sin 2x \sin z \cos y + \sin 2y \sin x \cos z + \sin 2z \sin y \cos x) - 0.2(\cos 2x \cos 2y + \cos 2y \cos 2z + \cos 2z \cos 2x) - 0.4(\cos x + \cos y + \cos z)$ |
| **Tubular / Skeletal Surfaces** | | |
| Tubular Gyroid | — | $10(\cos x \sin y + \cos y \sin z + \cos z \sin x) - 0.5(\cos 2x \cos 2y + \cos 2y \cos 2z + \cos 2z \cos 2x) - 14$ |
| Tubular D | — | $10(\sin x \sin y \sin z + \sin x \cos y \cos z + \cos x \sin y \cos z + \cos x \cos y \sin z) - 0.7(\cos 4x + \cos 4y + \cos 4z) - 11$ shifted by $\pi/4$ in each axis |
| Tubular P | — | $10(\cos x + \cos y + \cos z) - 5.1(\cos x \cos y + \cos y \cos z + \cos z \cos x) - 14.6$ |
| **"Difficult" Surfaces** | | |
| F Surface | F*, triplepanel | $\cos x \cos y \cos z$ |
| W Surface | W* | $(\cos 2x \cos y + \cos 2y \cos z + \cos 2z \cos x) - (\cos x \cos 2y + \cos y \cos 2z + \cos z \cos 2x)$ |

## 7.2. Blocks

The most reused blocks are the product of cosines (C : Triple cos) used in 12 functions, the sum of pairwise products (XY, YZ, and YZ) of double (12) and single (11) frequency cosines, and the sum of double (8) and single (7) frequency cosines (see Table 8). Many surfaces are essentially linear combinations of just 2-3 scalar multiples of these blocks, optionally with constants added or subtracted, which shift the 0-level of the isosurface. For instance, the K surface $=0.3E + 0.3F - 0.4G + 0.2$.

*Table 8: Functions from Table 7, analyzed by Claude*

| Block | Mathematical Expression | # | Used In |
|---|---|---|---|
| C : Triple cos | $\cos x \cos y \cos z$ | 12 | Diamond, FRD, FRD Prime, PN Surface, Y Surface, C(Y) Surface, PMY, CPMY, Neovius, S9 Surface, Double D, F Surface |
| I : 2ω pairwise cos | $\cos 2x \cos 2y + \cos 2y \cos 2z + \cos 2z \cos 2x$ | 12 | Lidinoid, FRD, FRD Prime, G Prime 1, G Prime 2, C(I2-Y**), Bionic Bone 1, Bionic Bone 2, Double Gyroid, I2-Y**, Split P, Tubular Gyroid |
| F : Pairwise cos | $\cos x \cos y + \cos y \cos z + \cos z \cos x$ | 11 | IWP, Fischer-Koch, OCTO, PN Surface, K Surface, Slotted-P, S3 Surface, S7 Surface, Double D 2, Double P, Tubular P |
| G : 2ω cos sum | $\cos 2x + \cos 2y + \cos 2z$ | 8 | IWP, Fischer-Koch, C(S) Surface, PN Surface, K Surface, Slotted-P, Double P, Tubular D |
| E : P-sum | $\cos x + \cos y + \cos z$ | 7 | P Surface, OCTO, PN Surface, K Surface, Neovius, Split P, Tubular P |
| A : Gyroid | $\sin x \cos y + \sin y \cos z + \sin z \cos x$ | 5 | Gyroid, C(G) Surface, Bionic Bone 1, Bionic Bone 2, Tubular Gyroid |
| H : Lidinoid triple | $\sin 2x \cos y \sin z + \sin 2y \cos z \sin x + \sin 2z \cos x \sin y$ | 5 | Lidinoid, G Prime 1, G Prime 2, Double Gyroid, Split P |
| D : Triple sin | $\sin x \sin y \sin z$ | 4 | Diamond, Y Surface, C(Y) Surface, Double D 2 |
| L : Mixed pairwise sin | $\sin 2x \sin y + \sin 2y \sin z + \sin 2z \sin x$ | 4 | Y Surface, C(Y) Surface, PMY, CPMY |

We can summarize the number of blocks shown in Table 8 even further if we consider that the sine function is just a cosine function phase-shifted by $\pi/2$, so $\sin(x)$ becomes $\cos(x-\pi/2)$. With this in mind, we can identify what we call "superblocks". The new winner, as the most common superblock, becomes the sum of cyclic pairwise $\cos(x+\varphi)\cos(y+\varphi)+\cos(y+\varphi)\cos(z+\varphi)+\cos(z+\varphi)\cos(x+\varphi), \varphi \in [0,\pi/2]$ with a total of 18 occurrences. And if we group even further into the format $\cos(Nx+\varphi)$, where N is any positive integer to account for different frequency combinations, this block appears in 13 additional functions, for a total of 31. The famous Gyroid could then be written as "cyclic pairwise, $\varphi = (0,\pi/2)$, N=1".

## 7.3. Additional Novel SFPH

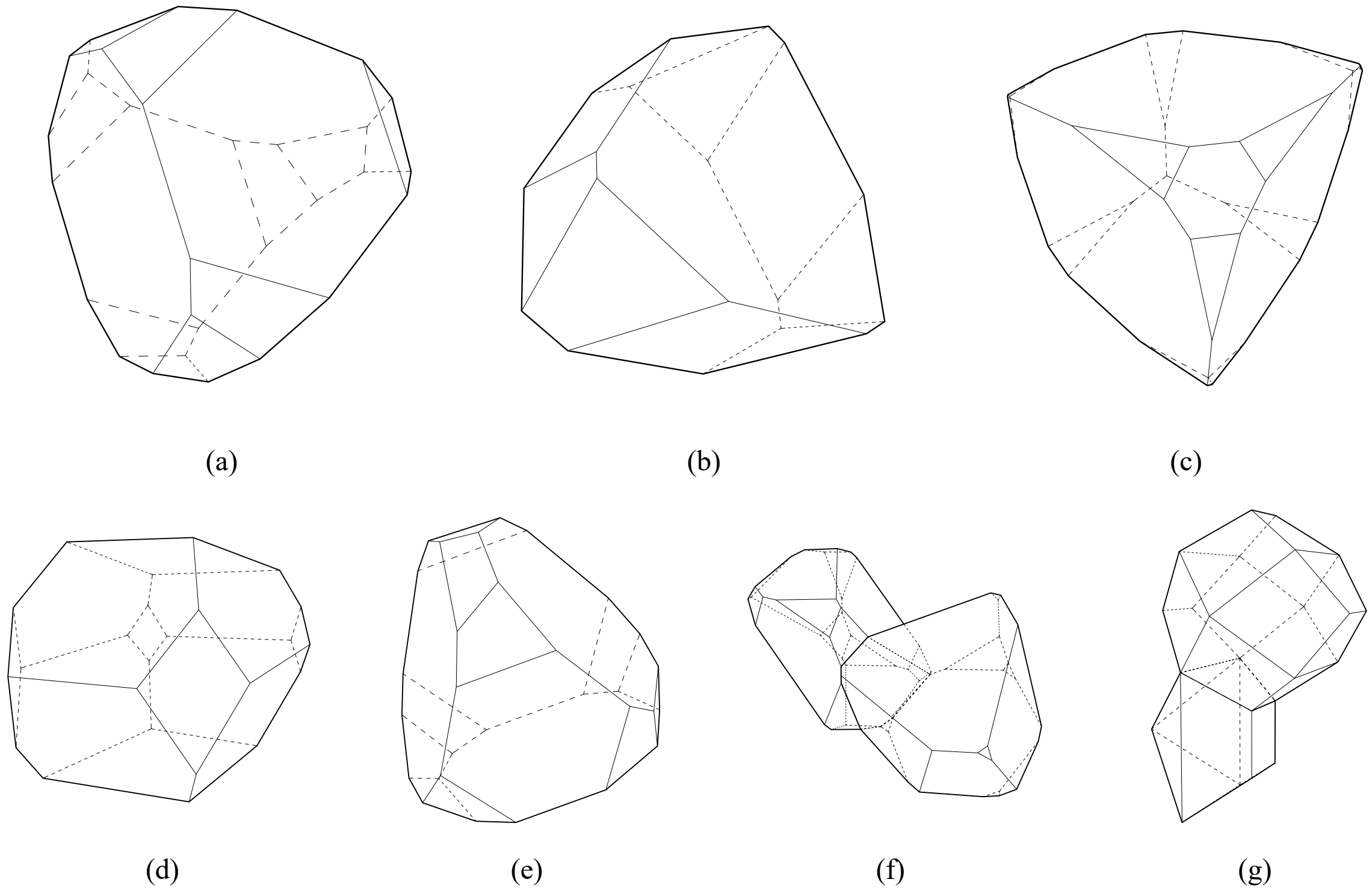

*Fig. 12: Cousins of the Josehedron, new types of SFPH: (a) Gyroid both, (b) Double Gyroid both, (c) FRP both, (d) Lidinoid max, (e) Split-P max, (f) Split-P both, (g) KP both*

It remains an assumption for now that the polyhedra shown in Fig. 12 are unique. They are space-filling, as they are created as Voronoi tessellations, but a study as profound as for the *Josehedron*, to investigate their uniqueness and special geometric properties, has not yet been conducted. One notable property is that the SFPH of the "Gyroid both" (Fig. 12 a) has among its faces two parallel regular hexagons. Due to the numerical approximation ininherent of the method, statistically significant similarity in face count and volume is used as a first indicator of potentially unique SFPH occurrence.